\documentclass[12pt]{article}
\usepackage{amsmath}
\usepackage{amsfonts}
\usepackage{amsthm}
\usepackage[top=3cm,left=3cm,right=3cm,bottom=3cm]{geometry}
\usepackage{graphicx}

\newtheorem{prop}{Proposition}
\theoremstyle{definition}
\DeclareMathOperator{\Tra}{Tr}
\newcommand{\p}{\partial}
\newcommand{\pfr}{\frac{\p f}{\p r}}
\newcommand{\Tr}[1]{\Tra{\left[#1\right]}}
\numberwithin{equation}{section}

\begin{document}

\begin{flushleft}

{\bf \Large Conditional symmetries and Riemann invariants for inhomogeneous hydrodynamic-type systems}\\[1cm]

A.M. Grundland$^{1,2}$ and B. Huard$^3$ \\
1. Centre de Recherches Math{\'e}matiques, Universit{\'e} de Montr{\'e}al, \\
C.P. 6128, Succ.\ Centre-ville, Montr{\'e}al, (QC) H3C 3J7, Canada\\
2. Universit{\'e} du Qu{\'e}bec, Trois-Rivi{\`e}res CP500 (QC) G9A 5H7, Canada\\ 
3. D{\'e}partement de math{\'e}matiques et de statistique, \\
C.P. 6128, Succ. Centre-ville, Montr{\'e}al, (QC) H3C 3J7, Canada
\end{flushleft}


\begin{abstract}
A new approach to the solution of quasilinear nonelliptic first-order systems of inhomogeneous PDEs in many dimensions is presented.  It is based on a version of the conditional symmetry and Riemann invariant methods.  We discuss in detail the necessary and sufficient conditions for the existence of rank-$2$ and rank-$3$ solutions expressible in terms of Riemann invariants.  We perform the analysis using the Cayley-Hamilton theorem for a certain algebraic system associated with the initial system.  The problem of finding such solutions has been reduced to expanding a set of trace conditions on wave vectors and their profiles which are expressible in terms of Riemann invariants.  A couple of theorems useful for the construction of such solutions are given.  These theoretical considerations are illustrated by the example of inhomogeneous equations of fluid dynamics which describe motion of an ideal fluid subjected to gravitational and Coriolis forces.  Several new rank-$2$ solutions are obtained.  Some physical interpretation of these results is given.

\end{abstract}


\section{Introduction}
\label{Sec-Introduction}


In the last three decades, a number of useful extensions of the classical Lie approach to group-invariant solutions of PDEs have generated a great deal of interest and activity in several fields of research, including mathematics, physics, chemistry and biology.  In particular, the method of partially invariant solutions \cite{olver-rosenau,ondich,ovsiannikov,zakharov}, the non-classical method \cite{bluman-kumei,levi-winternitz-nonclassical}, the conditional symmetry method \cite{grundland-huard-jnmp,grundland07-1,grundland-martina-rideau,rozdestvenskii-janenko,yanenko}, the weak symmetry method \cite{anderson-fels-torre,grundland-tempesta-winternitz} and the introduction of general "side conditions" or differential constraints \cite{olver-rosenau-weak} have been shown to play an essential role in several applications to nonlinear phenomena.  In particular, the method of differential constraints incorporates all known methods for determining particular solutions of PDEs.  They establish a direct connection between certain classes of solutions and a framework based on the theory of overdetermined nonlinear systems of PDEs.  In an attempt to understand certain classes of solutions expressible in terms of Riemann invariants, we have introduced a specific version of the conditional symmetry method \cite{grundland07-1}.  The basic idea was to analyze a Lie module of vector fields which are symmetries of an overdetermined system defined by the initial system of equations and certain first-order differential constraints.  It was shown \cite{grundland07-1} that this overdetermined system admits rank-$k$ solutions expressible in terms of Riemann invariants.  The method of conditional symmetry has led to a large number of new results.  

This paper is a follow-up of our previous investigations which were performed in \cite{conte-grundland-huard-jphysa-2009,grundland-huard-jnmp,grundland07-1,huard-non-homo}.  Here, we are using the Riemann invariants method for inhomogeneous nonelliptic quasilinear first-order systems of $l$ PDEs in $p = n+1$ independent variables of the form
\begin{equation}
\label{original-system}
u_t + A^i(u) u_{x^i} = B(u),
\end{equation}
where $x = (t,x^1, \ldots, x^n) = (x^0, \bar{x}) \in \mathbb{R}^{n+1}$ and $u = (u^1,\ldots,u^q) \in \mathbb{R}^q$ with the initial condition
$$x_0 = 0 \quad : \quad u(0, \bar{x}) = u_0(\bar{x}) \in \mathbb{R}^q.$$
As usual in mathematical physics, the space $X$ of independent variables $(x_0, \bar{x}) \in \mathbb{R}^p$ is called the physical space while the space of values of dependent variables $u = (u^1(x), \ldots, u^q(x))$ is denoted by $U \subset \mathbb{R}^q$ and is called the hodograph space.  The terms $u_t, u_{x^i}$ denote the first-order partial derivatives of $u$ with respect to $t$ and $x^i$, respectively.  The matrices $A^i(u)$ are $l \times q$ matrix functions of $u$ and $B(u)$ is an $l$-component vector.  Throughout this paper, we use the summation convention over repeated indices.  All our considerations are local. It suffices to search for solutions defined on a neighborhood of $x=0$.  The solutions $u=u(x)$ of the system (\ref{original-system}) are identified with their graphs which are $(n+1)$-dimensional manifolds in the cartesian product $X \times U$.

In our previous analysis of first-order systems of the form (\ref{original-system}) (see e.g. \cite{grundland07-1}), we considered homogeneous systems, i.e. systems with $B(u) = 0$, with coefficients depending only on the unknown functions $u$.  The methodological approach proposed in these works was based on some generalization of the Riemann invariants method.  A specific aspect of that approach is the presence of both algebraic and geometric points of view.  A conversion of systems of PDEs into algebraic form was made possible by representing the so called integral elements as linear combinations of simple elements \cite{grundland07-1} associated with those vectors fields which generate characteristic curves in the $X \times U$ space.  The introduction of those elements proved to be fruitful from the point of view of constructing the rank-$k$ solutions in closed form by means of the Cayley-Hamilton theorem applied directly to the algebraic form of the system.  Using a variant of the conditional symmetry method, we have shown that these solutions comprise among others multiple wave solutions as a superposition of two or more Riemann waves.  The results obtained for the homogeneous systems were so promising that it seemed worthwhile to try to extend this approach and check its effectiveness for the case of nonelliptic inhomogeneous partial differential first-order systems.  This is, in short, the objective of this paper.  


The paper is organized as follows. In Section 2, we investigate the group-invariance properties of rank-$2$ and rank-$3$ solutions for inhomogeneous systems (\ref{original-system}) and obtain necessary and sufficient conditions for their existence. Section 3 illustrates the construction of such solutions for the inhomogeneous Euler equations describing a $(3+1)$-dimensional fluid flow under the influence of gravitational and Coriolis effects.  Results and future perspectives are summarized in Section 4.


\section{Rank-$k$ solutions described by inhomogeneous systems of PDEs}


A version of the conditional symmetry method \cite{conte-grundland-huard-jphysa-2009,grundland-huard-jnmp,grundland07-1} has been developed recently for homogeneous nonelliptic systems (\ref{original-system}).  This method consists of supplementing the original system of PDEs (\ref{original-system}) with first-order differential constraints (DCs) for which a symmetry criterion for the given system of PDEs is identically satisfied.  It turns out that under certain circumstances, the so-augmented system of PDEs admits a larger class of Lie symmetries than the original system of PDEs (\ref{original-system}).  We now extend this method to inhomogeneous systems (\ref{original-system}) and reformulate the task of constructing rank-$k$ solutions expressible in terms of Riemann invariants in the language of the group analysis of differential equations. For this purpose, we require the solution $u$ of PDEs (\ref{original-system}) to be invariant under the family of commuting first-order differential operators
\begin{equation}
\label{vector-fields}
X_a = \xi^i_a(u) \frac{\p}{\p x^i}, \quad a= 1, \ldots, p-k, \quad k < p,
\end{equation}
defined on $X \times U$ and satisfying the conditions
\begin{equation}
\label{orthogonality}
\xi^i_a \lambda^s_i = 0, \quad s = 0, \ldots, k-1, \quad a = 1,\ldots,p-k.
\end{equation}
Here, the inhomogeneous wave vector $\lambda^0$ satisfies the rank condition
\begin{equation}
\label{Kronecker-Capella}
\mathrm{rank} \left( \lambda^0_i A^i ,  B(u) \right) = \mathrm{rank} \left( \lambda^0_i A^i \right),
\end{equation}
while the wave vectors $\lambda^0, \lambda^1, \ldots, \lambda^{k-1}$ are linearly independent vectors satisfying the dispersion relation for system (\ref{original-system})
\begin{equation*}
\label{Dispersion-relation}
\mathrm{rank} (\lambda^s_i A^i  ) < l, \quad s = 1,\ldots, k-1.
\end{equation*}
The group-invariant solutions of the system (\ref{original-system}) then consist of those functions $u=f(x)$ which satisfy both the initial system (\ref{original-system}) and a set of first-order differential constraints 
\begin{equation}
\label{first-order-DCs}
\xi^i_a \frac{\p u^{\alpha}}{\p x^i} = 0, \quad a=1,\ldots, p-k, \quad \alpha = 1,\ldots,q,
\end{equation}
ensuring that the characteristics of the vector fields $X_a$ are equal to zero.  In this case, if the vector function $u(x)$ is invariant under a set of $(p-k)$ vector fields $X_a$ for which the orthogonality property (\ref{orthogonality}) is satisfied, then the solution $u(x)$ of (\ref{original-system}) can be defined implicitly by the following set of relations between the variables $u^{\alpha}, x^i$ and $r^s$ : 
\begin{equation}
\label{rank-k-sol}
u = f(r^0, \ldots, r^{k-1}) , \quad r^s( x, u ) = \lambda^s_0({u}) t + \lambda^s_i({u}) x^i,  \quad s = 0,\ldots,k-1.
\end{equation}
Each function $r^s({x,u})$ is called the Riemann invariant associated with the vector $\lambda^s$.  

A vector field $X_a$ defined on $X \times U$ is said to be a conditional symmetry of system (\ref{original-system}) if it is tangent to $\mathbb{S} = \mathbb{S}_{\Delta} \cap \mathbb{S}_{Q}$, where $\mathbb{S}_{\Delta}$ and $\mathbb{S}_{Q}$ are submanifolds of the solution spaces defined by
\begin{equation*}
\begin{split}
&\mathbb{S}_{\Delta} = \left\{ \left(x,u^{(1)}\right) : u_t + A^i(u) u_{x^i} = B(u) \right\},\\
&\mathbb{S}_{Q} = \left\{ \left(x,u^{(1)}\right) : \xi^i_a u^{\alpha}_{x^i} = 0, \quad \alpha = 1,\ldots, q, \quad a = 1, \ldots, p-k \right\}.
\end{split}
\end{equation*}
An Abelian Lie algebra $L$ spanned by the vector fields $X_1,
\ldots, X_{p-k}$ is called a conditional symmetry algebra of the
original system (\ref{original-system}) if the following condition
\begin{equation*}
\label{cond-symmetry}
\mathrm{pr}^{(1)} X_a \left(u_t + A^i(u) u_{x^i} - B(u)\right)\Big|_{\mathbb{S}} =0, \quad a=1,\ldots, p-k, 
\end{equation*}
is satisfied, where $\mathrm{pr}^{(1)} X_a$ is the first prolongation of $X_a$.  

When studying solutions of type (\ref{rank-k-sol}), it is convenient from the computational point of view to write system (\ref{original-system}) in the form of a trace equation,
\begin{equation}
\label{original-system-trace}
\Tr{{\cal A}^{\mu}({u}) {\partial u}} = B^{\mu}({u}), \quad \mu = 1,\ldots, l,
\end{equation}
where ${\cal A}^{\mu}({u})$ are $p\times q$ matrix functions of ${u}$.  The Jacobian matrix of relations (\ref{rank-k-sol}) can be expressed in matrix form either as
\begin{equation}
\label{Jacobian-1}
\partial u = (u^{\alpha}_{x^i}) = \frac{\p f}{\p r} \left({\cal I}_k - \left(\eta_0 t + \eta_i x^i\right) \frac{\p f}{\p r} \right)^{-1} \lambda,
\end{equation} 
or as
\begin{equation}
\label{Jacobian-2}
\partial u =  \left({\cal I}_q - \frac{\p f}{\p r} \left(\eta_0 t + \eta_i x^i\right)  \right)^{-1} \frac{\p f}{\p r}\lambda,
\end{equation} 
where 
\begin{equation*}
\begin{split}
&\frac{\p f}{\p r} = \left( \frac{\p f^{\alpha}}{\p r^s} \right) \in \mathbb{R}^{q \times k}, \quad
\lambda = (\lambda^s_i) \in \mathbb{R}^{k\times p}, \quad \eta_i = \left(\frac{\p \lambda^s_i}{\p u^{\alpha}}\right), \quad i = 0, \ldots, n,
\end{split}
\end{equation*}
and ${\cal I}_k$ and ${\cal I}_q$ are the $k\times k$ and $q \times q$ identity matrices respectively.

A solution of the form (\ref{rank-k-sol}) is called a rank-$k$ solution if, in some open set of the origin $x=0$, we can express $u$ explicitly as a graph over the space of independent and dependent variables and the condition 
$$\mathrm{rank} (\partial u) = k$$ 
holds.

The condition of inversibility of the matrices appearing in equations (\ref{Jacobian-1}) and (\ref{Jacobian-2}) 
$$M_1 = {\cal I}_k - \left(\eta_0 t + \eta_i x^i\right) \frac{\p f}{\p r}, \quad M_2 = {\cal I}_q - \frac{\p f}{\p r} \left(\eta_0 t + \eta_i x^i\right),$$
restricts the domain of existence of rank-$k$ solutions.  However, since the inverse matrices are well defined at $x = 0$, there exists a neighborhood of the origin in which $\det(M_i) \neq 0$, $i=1,2$, allowing us to look for rank-$k$ solutions locally parametrized by equations (\ref{rank-k-sol}).
It should also be noted that because of the Weinstein-Aronzjain determinant relation
\begin{equation}
\label{Weinstein-Aronzjain}
\det{\left({\cal I}_k - P Q\right)} = \det{\left({\cal I}_q - Q P\right)}, \quad P \in \mathbb{R}^{k\times q}, \quad Q \in \mathbb{R}^{q \times k},
\end{equation}
which can be derived from the relation
\begin{equation*}
\det{ \left[ \left( \begin{array}{cc} {\cal I}_k & P \\ Q & {\cal I}_q \end{array} \right) \left( \begin{array}{cc} {\cal I}_k & 0 \\ -Q & {\cal I}_q \end{array} \right) \right] } =  \det{ \left[ \left( \begin{array}{cc} {\cal I}_k & 0 \\ -Q & {\cal I}_q \end{array} \right)\left( \begin{array}{cc} {\cal I}_k & P \\ Q & {\cal I}_q \end{array} \right)  \right] } ,
\end{equation*}
we have that $\det{(M_1)} = \det{(M_2)}$. Hence, the regions of validity for the expressions (\ref{Jacobian-1}) and (\ref{Jacobian-2}) are the same.  Morever, (\ref{Weinstein-Aronzjain}) implies that $\det{(M_1)}$ is a polynomial of order $s = \min{(k,q)}$.  Inserting (\ref{Jacobian-1}) into the original system (\ref{original-system-trace}), we obtain the result
\begin{equation*}
\Tr{{\cal A}^{\mu}(u) \left({\cal I}_k - \left(\eta_0 t + \eta_i x^i\right) \frac{\p f}{\p r} \right)^{-1} \lambda } = B^{\mu}(u),
\end{equation*}
or, upon multiplication by $\det{(M_1)}$, 
\begin{equation}
\label{Traces-poly}
\Tr{{\cal A}^{\mu}(u) \mathrm{adj}\left({\cal I}_k - \left(\eta_0 t + \eta_i x^i\right) \frac{\p f}{\p r} \right) \lambda } - \det{\left[{\cal I}_k - \left(\eta_0 t + \eta_i x^i\right) \frac{\p f}{\p r} \right]}B^{\mu}(u) = 0,
\end{equation}
which is a polynomial expression of degree $s$ in the independent variables $(t,x^i)$.  For the system (\ref{Traces-poly}) to be expressed only in terms of Riemann invariants, we must require that every coefficient of $(t, x^i)$ of this polynomial vanish identically.  We then consider equation (\ref{Traces-poly}) and all its partial derivatives up to order $s$ with respect to the independent variables $(t,x^i)$ taken at the origin and set the resulting expressions to zero.
The partial derivatives of the trace equation (\ref{Traces-poly}) are obtained by making use of the Jacobi equations for the derivative of the determinant of any square matrix $M=M(\xi)$ depending on a parameter $\xi$,
\begin{equation}
\label{rel-derivatives}
\begin{split}
&\frac{\p}{\p \xi} \det{\left[M\right]} = \Tr{\mathrm{adj}[M] \frac{\p M}{\p\xi}},\\
&\frac{\p}{\p \xi} \mathrm{adj}[M] = \left(\Tr{\mathrm{adj}[M] \frac{\p M}{\p \xi}} {\cal I} - \mathrm{adj}[M] \frac{\p M}{\p \xi} \right) M^{-1}.
\end{split}
\end{equation}
A conditionally invariant solution of (\ref{original-system}) under an Abelian algebra $L$ spanned by vector fields $X_1, \ldots, X_{p-k}$ of the form (\ref{vector-fields}) is then obtained by solving the overdetermined system
\begin{equation}
\label{overdetermined-system}
\Delta' : \begin{cases} \displaystyle \Tr{{\cal A}^{\mu}(u) \mathrm{adj}\left({\cal I}_k - \left(\eta_0 t + \eta_i x^i\right) \frac{\p f}{\p r} \right) \lambda } - \det{\left[{\cal I}_k - \left(\eta_0 t + \eta_i x^i\right) \frac{\p f}{\p r} \right]}B^{\mu}(u) = 0, \\  
\,\xi^i_a u^{\alpha}_{x^i} = 0, \quad a=1,\ldots, p-k, \quad \alpha = 1, \ldots, q.
\end{cases}
\end{equation}
Therefore, in the case $k=2$, we can devise the following proposition.
\begin{prop}
Consider a fixed set of linearly-independent wave vectors $\lambda^0$ and $\lambda^1$ associated with a nondegenerate quasilinear nonelliptic first-order system of PDEs (\ref{original-system}) in $p$ independent variables and $q$ dependent variables.  This system admits a $(p-2)$-dimensional conditional symmetry algebra $L$ if and only if $(p-2)$ linearly independent vector fields
$$X_a = \xi^i_a (u) \frac{\p}{\p x^i}, \quad \lambda^s_i \xi^i_a = 0, \quad a=1,\ldots, p-2, \quad s = 0,1,$$
satisfy the conditions
\begin{subequations}
\label{rank-2-trace-eqs}
\begin{eqnarray}
\label{rank-2-trace-eqs-a}
&&\Tr{{\cal A}^{\mu} \frac{\p f}{\p r} \lambda } = B^{\mu}, \quad \mu = 1,\ldots, l,\\
\label{rank-2-trace-eqs-b}
&&\Tr{{\cal A}^{\mu} \frac{\p f}{\p r} \eta_i \frac{\p f}{\p r} \lambda} = 0, \quad i=0,\ldots, n,\\
\label{rank-2-trace-eqs-c}
&&\Tr{{\cal A}^{\mu} \frac{\p f}{\p r} \left( \eta_i \frac{\p f}{\p r} \eta_j + \eta_j \frac{\p f}{\p r} \eta_i \right) \frac{\p f}{\p r} \lambda } = 0, \quad i \neq j = 0,\ldots,n,\\
\label{rank-2-trace-eqs-d}
&&\det{\left[ \eta_i \frac{\p f}{\p r} \right]} = 0,  \quad \eta_i = \left(\frac{\p \lambda^s_i}{\p u^{\alpha}} \right) \in \mathbb{R}^{2 \times q}.
\end{eqnarray}
\end{subequations}
and such that $\mathrm{rank}{\left(\p f / \p r \right)} = 2 $ on some neighborhood of a point $(x_0,u_0) \in \mathbb{S}$. The solution of (\ref{original-system}) which is invariant under the Lie algebra $L$ is precisely a rank-$2$ solution of the form (\ref{rank-k-sol}).
\end{prop}
\begin{proof}
It was shown in \cite{grundland07-1} that in the new coordinates $\mathbb{R}^p \times \mathbb{R}^q$
\begin{equation*}
\label{rank-$k$-coord}
\begin{split}
&\bar{x}^1 = r^1(x,u), \bar{x}^2 = r^2(x,u), \bar{x}^{3} = x^{3},\, \ldots,\, \bar{x}^p = x^p, \bar{u}^1 = u^1,\, \ldots,\, \bar{u}^q = u^q, \end{split}
\end{equation*}
the vector fields $X_a$ adopt the rectified form
$$X_a = \frac{\partial}{\partial_{\bar{x}^{a}}}, \quad a = 3, \ldots, p.$$
Hence, the symmetry criterion for $G$ to be the symmetry group of
the overdetermined system (\ref{overdetermined-system}) requires that the vector fields
$X_a$ of $G$ satisfy
\begin{equation*}
  X_a(\Delta') = 0,
\end{equation*}
whenever equations (\ref{overdetermined-system}) hold.  Thus the symmetry criterion
applied to the invariance conditions (\ref{first-order-DCs}) is
identically equal to zero.  After applying the symmetry criterion to the system (\ref{Traces-poly}) in new coordinates and taking into account the conditions (\ref{rank-2-trace-eqs}), we obtain the equations which are identically satisfied.

The converse is also true.  The requirement that the system (\ref{original-system}) be nondegenerate means that it is locally solvable and is of maximal rank at every point $(x_0, u_0^{(1)}) \in \mathbb{S}$.  Therefore \cite{olver-2000}, the infinitesimal symmetry criterion is a necessary and sufficient condition for the existence of the symmetry group $G$ of the overdetermined system (\ref{overdetermined-system}).  Using relations (\ref{rel-derivatives}), we show that equations (\ref{rank-2-trace-eqs}) must hold.  Equations (\ref{rank-2-trace-eqs-a}) and (\ref{rank-2-trace-eqs-b}) are easily obtained by considering the system (\ref{Traces-poly}) and all its first-order partial derivatives with respect to $t$ and $x^i$.  The second-order derivatives with respect to $x^i$ and $x^j$ provide us with the condition
\begin{equation*}
\label{rank-2-trace-eqs-e}
\Tr{{\cal A}^{\mu} \frac{\p f}{\p r} \left( \eta_i \frac{\p f}{\p r} \eta_j + \eta_j \frac{\p f}{\p r} \eta_i \right) \frac{\p f}{\p r} \lambda } = 0, \quad i,j = 0,\ldots,n,
\end{equation*}
which reduces when $i=j$ to
\begin{equation}
\label{rank-2-trace-eqs-f}
\Tr{{\cal A}^{\mu} \frac{\p f}{\p r} \eta_i \frac{\p f}{\p r} \eta_i \frac{\p f}{\p r} \lambda } = 0, \quad i = 0,\ldots,n.
\end{equation}
Considering equations (\ref{rank-2-trace-eqs-b}), we can write (\ref{rank-2-trace-eqs-f}) as
\begin{eqnarray}
\hspace{-2.5cm}
\Tr{{\cal A}^{\mu} \frac{\p f}{\p r} \eta_i \frac{\p f}{\p r} \eta_i \frac{\p f}{\p r} \lambda } 
& = & \Tr{{\cal A}^{\mu} \frac{\p f}{\p r} \eta_i \frac{\p f}{\p r} \eta_i \frac{\p f}{\p r} \lambda} -\Tr{ \eta_i \frac{\p f}{\p r}}  \Tr{{\cal A}^{\mu} \frac{\p f}{\p r} \eta_i \frac{\p f}{\p r} \lambda} \nonumber \\
\label{rank-2-trace-eqs-g}
& = & \Tr{{\cal A}^{\mu} \frac{\p f}{\p r} \eta_i \frac{\p f}{\p r} \left( \eta_i \frac{\p f}{\p r} - \Tr{ \eta_i \frac{\p f}{\p r} } {\cal I}_2 \right) \lambda }.
\end{eqnarray}
According to the Cayley-Hamilton theorem, each matrix $\left(\eta_i \frac{\p f}{\p r}\right) \in \mathbb{R}^{2\times 2}$ satifies
\begin{equation*}
\left(\eta_i \frac{\p f}{\p r} \right)^2 - \Tr{ \eta_i\frac{\p f}{\p r} } \left( \eta_i\frac{\p f}{\p r} \right) = - \det{\left[ \eta_i\frac{\p f}{\p r} \right]} {\cal I}_2 
\end{equation*}
allowing us to write (\ref{rank-2-trace-eqs-g}) as
\begin{equation*}
\Tr{{\cal A}^{\mu} \frac{\p f}{\p r} \eta_i \frac{\p f}{\p r} \eta_i \frac{\p f}{\p r} \lambda } = \det{\left[\eta_i \frac{\p f}{\p r} \right]} \Tr{{\cal A}^{\mu} \frac{\p f}{\p r} \lambda } = B^{\mu} \det{\left[\eta_i \frac{\p f}{\p r} \right]},
\end{equation*}
by (\ref{rank-2-trace-eqs-a}).  As a consequence of the assumption $B^{\mu} \not \equiv 0$, we must have that
$$\det{\left[\eta_i \frac{\p f}{\p r} \right]} = 0.$$  That ends the proof, since the solutions of the overdetermined system (\ref{overdetermined-system}) are invariant under the algebra $L$ generated by the $(p-k)$ vector fields $X_1, \ldots, X_{p-2}$.  The invariants of the group $G$ of such vector fields are provided by the functions $\{r^0, r^1, u^1, \ldots, u^q\}$.  So the general rank-$k$ solution of (\ref{original-system}) takes the form (\ref{rank-k-sol}).

\end{proof}
Note that in the case $q = k = 2$, equations (\ref{rank-2-trace-eqs-d}) imply that
\begin{equation*}
\det{\left[\eta_i \frac{\p f}{\p r} \right]} = \det{\left[\eta_i\right]} \det{\left[\frac{\p f}{\p r}\right]} = 0.
\end{equation*}
When $\det{\left[\frac{\p f}{\p r}\right]} = 0$, the considered solution reduces to a rank-$1$ solution. Consequently, when $q=2$, rank-$2$ solutions exist only when $\det{\left[\eta_i\right]} = 0$ for $i=0,\ldots,n$. Note that in every case, equations (\ref{rank-2-trace-eqs-d}) can be satisfied, for example, by choosing one of the vectors $\lambda^s$ to be constant.

Analogously, we obtain the following conditions in the case $k=3$.
\begin{prop}
Consider a fixed set of linearly-independent wave vectors $\lambda^0$, $\lambda^1$ and $\lambda^2$ associated with a nondegenerate quasilinear nonelliptic first-order system of PDEs (\ref{original-system}) in $p$ independent variables and $q$ dependent variables.  This system admits a $(p-3)$-dimensional conditional symmetry algebra $L$ if and only if $(p-3)$ linearly independent vector fields
$$X_a = \xi^i_a (u) \frac{\p}{\p x^i}, \quad \lambda^s_i \xi^i_a = 0, \quad a=1,\ldots, p-3, \quad s = 0,1,2,$$
satisfy the conditions
\begin{subequations}
\label{rank-3-trace-eqs}
\begin{eqnarray}
\label{rank-3-trace-eqs-a}
&&\hspace{-2cm} \Tr{{\cal A}^{\mu} \frac{\p f}{\p r} \lambda } = B^{\mu}, \quad \mu = 1,\ldots, l,\\
\label{rank-3-trace-eqs-b}
&&\hspace{-2cm}\Tr{{\cal A}^{\mu} \frac{\p f}{\p r} \eta_{i_1} \frac{\p f}{\p r} \lambda} = 0, \quad i_1=0,\ldots, n,\\
\label{rank-3-trace-eqs-c}
&&\hspace{-2cm}\Tr{{\cal A}^{\mu} \frac{\p f}{\p r} \left( \eta_{i_1} \frac{\p f}{\p r} \eta_{i_2} + \eta_{i_2} \frac{\p f}{\p r} \eta_{i_1} \right) \frac{\p f}{\p r} \lambda } = 0, \quad i_1, i_2 = 0,\ldots,n,\\
\label{rank-3-trace-eqs-d}
&&\hspace{-2cm}\Tra \left[{\cal A}^{\mu} \frac{\p f}{\p r} \left( \eta_{i_1} \frac{\p f}{\p r} \eta_{i_2} \frac{\p f}{\p r} \eta_{i_3} + \eta_{i_1} \frac{\p f}{\p r} \eta_{i_3} \frac{\p f}{\p r} \eta_{i_2}  + \eta_{i_2} \frac{\p f}{\p r} \eta_{i_1} \frac{\p f}{\p r} \eta_{i_3}  \right. \right. \nonumber \\
&&\hspace{-2cm} \left. \left. \qquad + \eta_{i_2} \frac{\p f}{\p r} \eta_{i_3} \frac{\p f}{\p r} \eta_{i_1}  + \eta_{i_3} \frac{\p f}{\p r} \eta_{i_1} \frac{\p f}{\p r} \eta_{i_2} + \eta_{i_3} \frac{\p f}{\p r} \eta_{i_2} \frac{\p f}{\p r} \eta_{i_1}  \right) \frac{\p f}{\p r} \lambda \right]  = 0, \\
\label{rank-3-trace-eqs-e}
&&\hspace{-2cm}\det{\left[ \eta_i \frac{\p f}{\p r} \right]} = 0,  \quad \eta_i = \left(\frac{\p \lambda^s_i}{\p u^{\alpha}} \right) \in \mathbb{R}^{3 \times q},
\end{eqnarray}
\end{subequations}
and such that $\mathrm{rank}{\left(\p f / \p r \right)} = 3$ on some neighborhood of a point $(x_0,u_0) \in \mathbb{S}$, where the indices $i_1,i_2,i_3$ take the values $0,\ldots,n$ and are not all equal.
\end{prop}
\begin{proof}
The proof for sufficiency follows the same steps as in the case $k=2$.  We now show the necessity of equations (\ref{rank-3-trace-eqs}). As previously, equations (\ref{rank-3-trace-eqs}a-d) represent the first, second and third-order partial derivatives of expression (\ref{Traces-poly}) taken at the origin, respectively.  When $i_1=i_2=i_3 = i$, equations (\ref{rank-3-trace-eqs-d}) reduce to
\begin{equation*}
\Tr{{\cal A}^{\mu} \frac{\p f}{\p r} \eta_i \frac{\p f}{\p r} \eta_i \frac{\p f}{\p r} \eta_i \frac{\p f}{\p r} \lambda } = 0. 
\end{equation*}
For any $3$ by $3$ matrix $M$, the Leverrier-Faddeev algorithm and the Cayley-Hamilton theorem allow us to write
\begin{equation}
\label{Cayley-Hamilton-3}
M^3 - \Tr{M} M^2 + \frac{1}{2} \left( \Tr{M}^2 - \Tr{M^2} \right) M - \det[M] \mathcal{I}_3 = 0.
\end{equation}
Hence, taking into account equations (\ref{rank-3-trace-eqs}) and (\ref{Cayley-Hamilton-3}), we get
\begin{eqnarray*}
& &\hspace{-2.5cm}\Tr{{\cal A}^{\mu} \frac{\p f}{\p r} \eta_i \frac{\p f}{\p r} \eta_i \frac{\p f}{\p r} \eta_i \frac{\p f}{\p r} \lambda } - \Tr{\eta_i \pfr } \Tr{{\cal A}^{\mu} \pfr \eta_i \pfr \eta_i \pfr \lambda} \\
& &\hspace{-3cm} \qquad = \Tr{ {\cal A}^{\mu} \pfr \left(\eta_i \pfr \right)^2 \left( \eta_i\pfr - \Tr{\eta_i \pfr } \mathcal{I}_3\right) \lambda} \\
& &\hspace{-3cm} \qquad = \Tr{ {\cal A}^{\mu} \pfr \left( -\frac{1}{2} \left( \Tr{\eta_i \pfr }^2 - \Tr{\left(\eta_i \pfr \right)^2} \right)\eta_i \pfr + \det{\left[\eta_i \pfr \right]} \mathcal{I}_3 \right) \lambda}\\
& &\hspace{-3cm} \qquad = -\frac{1}{2} \left( \Tr{\eta_i \pfr }^2 - \Tr{\left(\eta_i \pfr\right)^2} \right) \Tr{ {\cal A}^{\mu} \pfr \eta_i \pfr \lambda} + \det{\left[\eta_i \pfr \right]} \Tr{{\cal A}^{\mu} \pfr \lambda}\\
& &\hspace{-3cm} \qquad = B^{\mu} \det{\left[\eta_i \pfr \right]},
\end{eqnarray*}
which is equal to zero if and only if the relations 
$$\det{\left[\eta_i \pfr \right]} = 0, \quad i=0,\ldots, n$$
hold.
Again, the solutions of the overdetermined system (\ref{overdetermined-system}) are invariant under the algebra of vector fields $X_1, \ldots, X_{p-3}$ defined by (\ref{vector-fields}), (\ref{orthogonality}) and (\ref{rank-3-trace-eqs}). The invariants of the group $G$ of such vector fields are provided by the functions $\{r^0, r^1, r^2, u^1,\ldots, u^q\}$, therefore the rank-$k$ solution of (\ref{original-system}) is of the form (\ref{rank-k-sol}).
\end{proof}


\section{Applications in fluid dynamics}


Now we present some examples which illustrate the theoretical considerations presented in Section 2.  We consider classical equations of hydrodynamics describing a motion in a fluid medium when the gravitational force $\vec{g}$ and Coriolis force $\Omega \times \vec{v}$ occur.  We restrict ourselves to the equations of a one-component nonviscous fluid flow.  Under these assumptions, our equations are of the type (\ref{original-system}).  The matrix form of these equations of hydrodynamics in a noninertial coordinates system is
\begin{equation}
\label{Euler-rotational}
A^0(u) u_t + A^1(u) u_x + A^2(u) u_y + A^3(u) u_z = B(u),
\end{equation}
where
\begin{equation*}
\begin{split}
&A^0(u) = 
\left(
\begin{array}{ccccc}
1 & 0 & 0 & 0 & 0 \\
0 & 1 & 0 & 0 & 0 \\
0 & 0 & 1 & 0 & 0 \\
0 & 0 & 0 & 1 & 0 \\
0 & 0 & 0 & -\kappa p / \rho & 1
\end{array}
\right), \quad
A^i(u) = 
\left(
\begin{array}{ccccc}
v_i & 0 & 0 & 0 & \delta_{i1}/\rho \\
0 & v_i & 0 & 0 & \delta_{i2}/\rho \\
0 & 0 & v_i & 0 & \delta_{i3}/\rho \\
0 & 0 & 0 & v_i & 0 \\
0 & 0 & 0 & -\kappa p v_i / \rho & v_i
\end{array}
\right), \quad i=1,2,3,\\
&B(u) = \left(\begin{array}{c}
g_1 - (\Omega_2 v_3 - \Omega_3 v_2) \\
g_2 - (\Omega_3 v_1 - \Omega_1 v_3) \\
g_3 - (\Omega_1 v_2 - \Omega_2 v_1) \\
0 \\
0 
\end{array}
\right), \quad u = (\vec{v}, \rho, p) \in \mathbb{R}^5, \quad \delta_{ij} = \begin{cases} 1 & i=j \\ 0 & i\neq j \end{cases}.
\end{split}
\end{equation*}
Here we treat the physical space $X \subset \mathbb{R}^4$ as classical space-time, each of its points having coordinates $(t,\vec{x}) = (t,x,y,z)$ and the space of unknown functions (i.e. the hodograph space) $U \subset \mathbb{R}^5$ having coordinates $(\vec{v},\rho,p)$.  The vector $\vec{\Omega}$ denotes the angular velocity while $\kappa$ is related to the adiabatic exponent $\gamma$ of the fluid by the relation $\kappa = 2(\gamma - 1)^{-1}$.

For the homogeneous system (\ref{original-system}), each wave vector $\lambda$ can be written in the form $\lambda = (\lambda_0, \vec{\lambda})$, where $\vec{\lambda} = (\lambda_1, \lambda_2, \lambda_3)$ denotes a direction of wave propagation and the eigenvalue $\lambda_0$ of the matrix $(\lambda_i A^i)$ is a phase velocity of the considered wave.  The dispersion relation for the homogeneous hydrodynamic system of equations (\ref{Euler-rotational}) takes the form
\begin{equation}
\label{Euler-dispersion-relation}
(\lambda_0 + \vec{v} \cdot \vec{\lambda})^3 \left[(\lambda_0 + \vec{\lambda}\cdot \vec{v})^2 - \frac{\kappa p}{\rho} |\vec{\lambda}|^2 \right] = 0.
\end{equation}
Solving equation (\ref{Euler-dispersion-relation}), we obtain two types of wave vectors, namely
\begin{equation*}
\begin{split}
&{\bf i) }\quad  \text{Entropic} : \lambda^E = \left( - \vec{\lambda} \cdot \vec{v}, \vec{\lambda} \right),\\
&{\bf ii) } \quad \text{Acoustic} : \lambda^{A_{\varepsilon}} = \left(\varepsilon \left(\frac{\kappa p}{\rho}\right)^{1/2} |\vec{\lambda}| - \vec{v} \cdot \vec{\lambda}, \vec{\lambda} \right), \quad \varepsilon = \pm 1.\\
\end{split}
\end{equation*}
Algebraic equations defining inhomogeneous wave vectors which satisfy equation (\ref{Kronecker-Capella}) are of the form
\begin{equation*}
\begin{split}
&{\bf i)} \quad \text{Entropic} : \lambda_{E^0} = \left(- \vec{v} \cdot \vec{g}, \vec{g} - \vec{\Omega} \times \vec{v}\right),\\
&{\bf ii)} \quad \text{Acoustic} : \lambda_{A_{\varepsilon}^0} = \left(\varepsilon \left(\frac{\kappa p}{\rho}\right)^{1/2} |\vec{\lambda}| - \vec{v} \cdot \vec{\lambda}, \vec{\lambda} \right), \quad \varepsilon = \pm 1\\
&{\bf iii)} \quad \text{Hydrodynamic} : \lambda_{H^0} = \left( \lambda_0 , \vec{\lambda} \right), \quad \lambda_0 + \vec{v} \cdot \vec{\lambda} \neq 0, \left(\frac{\kappa p}{\rho}\right)^{1/2} |\vec{\lambda}|.
\end{split}
\end{equation*}
Several classes of wave solutions of the hydrodynamic system (\ref{Euler-rotational}) have been obtained via the generalized method of characteristics \cite{grundland-non-homo-1974,jeffrey}.  Applying the CSM to this system allows us to compare the effectiveness of these two approaches.  Here we restrict ourselves to the consideration of superpositions of waves which are admissible by the inhomogeneous system (\ref{Euler-rotational}).  We are interested only in rank-$2$ solutions which can be written in Riemann invariants.  The results of such superpositions are set in Table \ref{Table-pm}, where the signs (+) and (-) indicate when superposition does and does not occur, respectively.  We will now show for illustration some physically interesting solutions from among the ones obtained.

\begin{table}[h]
\begin{center}
$\begin{array}{l|c|c}
        &       E       &       A_{\varepsilon}       \\\hline
 E^0    &      \hspace{2cm} +\hspace{2cm}       &     \hspace{2cm}  + \hspace{2cm}      \\\hline
 A^0_{\varepsilon}    &       +       &       -       \\\hline
 H^0    &       +       &       +       \\
\end{array}$
\caption{The existence (+) or absence (-) of nonlinear superposition of waves admitted by the system (\ref{Euler-rotational}).}
\label{Table-pm}
\end{center}
\end{table}

Different possibilities of existence of solutions in Riemann invariants for different combinations of wave vectors (homogeneous and inhomogeneous) are denoted by subsystems $E^0 E$, $E^0 A_{\varepsilon}$, $A^0_{\varepsilon} E$, $A^0_{\varepsilon} A_{\varepsilon}$, $H^0 E$, $H^0 A_{\varepsilon}$.  Moreover, we denote by $r^0$ a Riemann invariant associated with wave vectors $\lambda^0_E, \lambda^0_{A_{\varepsilon}}, \lambda^0_H$ and by $r^1$ a Riemann invariant associated with $\lambda^E, \lambda^{A_{\varepsilon}}$.

Now we present some examples which illustrate the theoretical considerations presented in Section 2.

{\bf Subsystem $\{E^0, E\}$ : } 
We look for solutions of the form (\ref{rank-k-sol}) defined by
\begin{eqnarray}
&&\hspace{-2cm}\vec{v}=\vec{v}(r^0,r^1), \quad \rho = \rho(r^0,r^1), \quad p = p(r^0,r^1), \nonumber\\
&&\hspace{-2cm}r^0 = - \left(\vec{v}\cdot\vec{g}\right) t + (g_1 + \Omega_3 v_2 - \Omega_2 v_3) x + (g_2 + \Omega_1 v_3 - \Omega_3 v_1) y + (g_3 + \Omega_2 v_1 - \Omega_1 v_2) z, \nonumber\\
&&\hspace{-2cm}r^1 = - \left(\vec{\lambda}^1 \cdot \vec{v} \right) t + \lambda^1_1 x + \lambda^1_2 y + \lambda^1_3 z, \quad \vec{\lambda}^1 = ( \lambda^1_1, \lambda^1_2, \lambda^1_3),
\end{eqnarray}
where the $\lambda^1_j$ are allowed to depend on the unknown functions.  To satisfy equations (\ref{rank-2-trace-eqs-d}), we require that the coefficients of $t,x,y,z$ in $r^0$ be constant, say $r^0 = c_0 t + c_1 x + c_2 y + c_3 z$. To allow for a rank-2 solution, the following relations must then hold
\begin{equation*}
\vec{g} \cdot \vec{\Omega} = 0, \quad v_1 = \frac{g_2 - c_2 + \Omega_1 v_3}{\Omega_3}, \quad v_2 = \frac{c_1 - g_1 + \Omega_2 v_3}{\Omega_3}, 
\end{equation*}
implying that this type of solution exists only when the rotation axis of the system is perpendicular to the constant gravitational force.  Solving (\ref{rank-2-trace-eqs}), we obtain the following solution
\begin{equation}
\label{E0E1-sol}
\begin{split}
&v_1 = \frac{g_2}{\Omega_3} - \frac{c_0}{g_1} - \frac{c_0 \Omega_2 g_2}{g_1 g_3 \Omega_3} + \frac{\Omega_1}{\Omega_3} v_3(r^0,r^1), \quad
v_2 = \frac{c_0 \Omega_2}{g_3 \Omega_3} - \frac{g_1}{\Omega_3} + \frac{\Omega_2}{\Omega_3} v_3(r^0,r^1),\\
&\rho = p_{r^0}, \quad p = p(r^0),
\end{split}
\end{equation}
where $p(r^0), v_3(r^0,r^1)$ are arbitrary functions of the Riemann invariants 
\begin{equation}
\label{E0E1-RI}
\begin{split}
&r^0 = \frac{c_0}{g_3} \left(g_3 t + \Omega_2 x - \Omega_1 y\right), \quad \vec{g} \cdot \vec{\Omega} = 0,\\
&r^1 =  \left( \frac{\lambda^1_2 g_1 - \lambda^1_1 g_2}{\Omega_3} + \frac{c_0 \lambda^1_3}{g_3} \right) t + \lambda^1_1 x + \lambda^1_2 y + \lambda^1_3 z,
\end{split}
\end{equation}
and $\vec{\lambda}^1$ is a constant vector satisfying $\vec{\lambda}^1 \cdot \vec{\Omega} = 0$.  The solution defined by (\ref{E0E1-sol}) and (\ref{E0E1-RI}) represents a double traveling wave with constant velocities.

\begin{figure}[h!]
\centering \includegraphics[width=6cm]{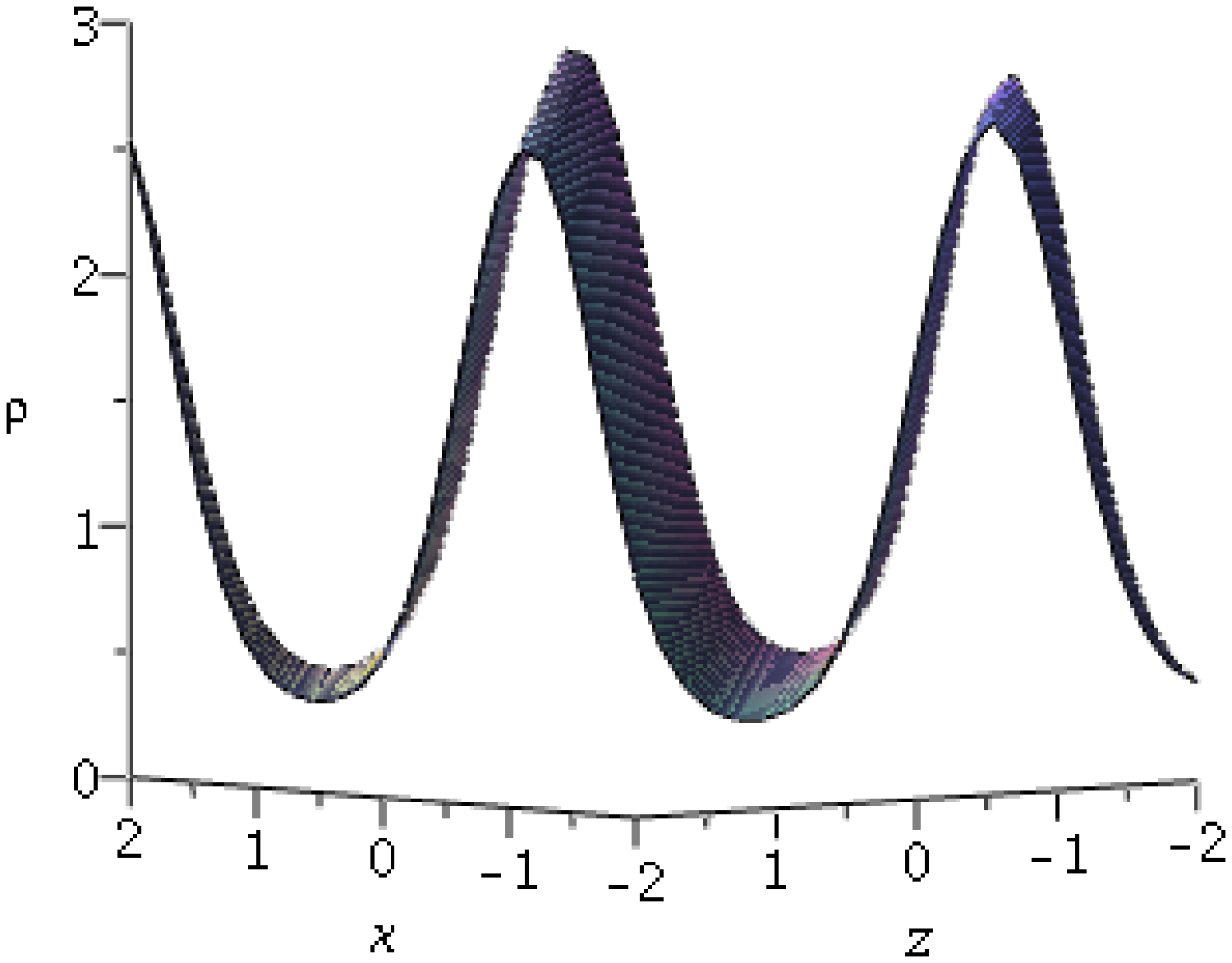} \includegraphics[width=6cm]{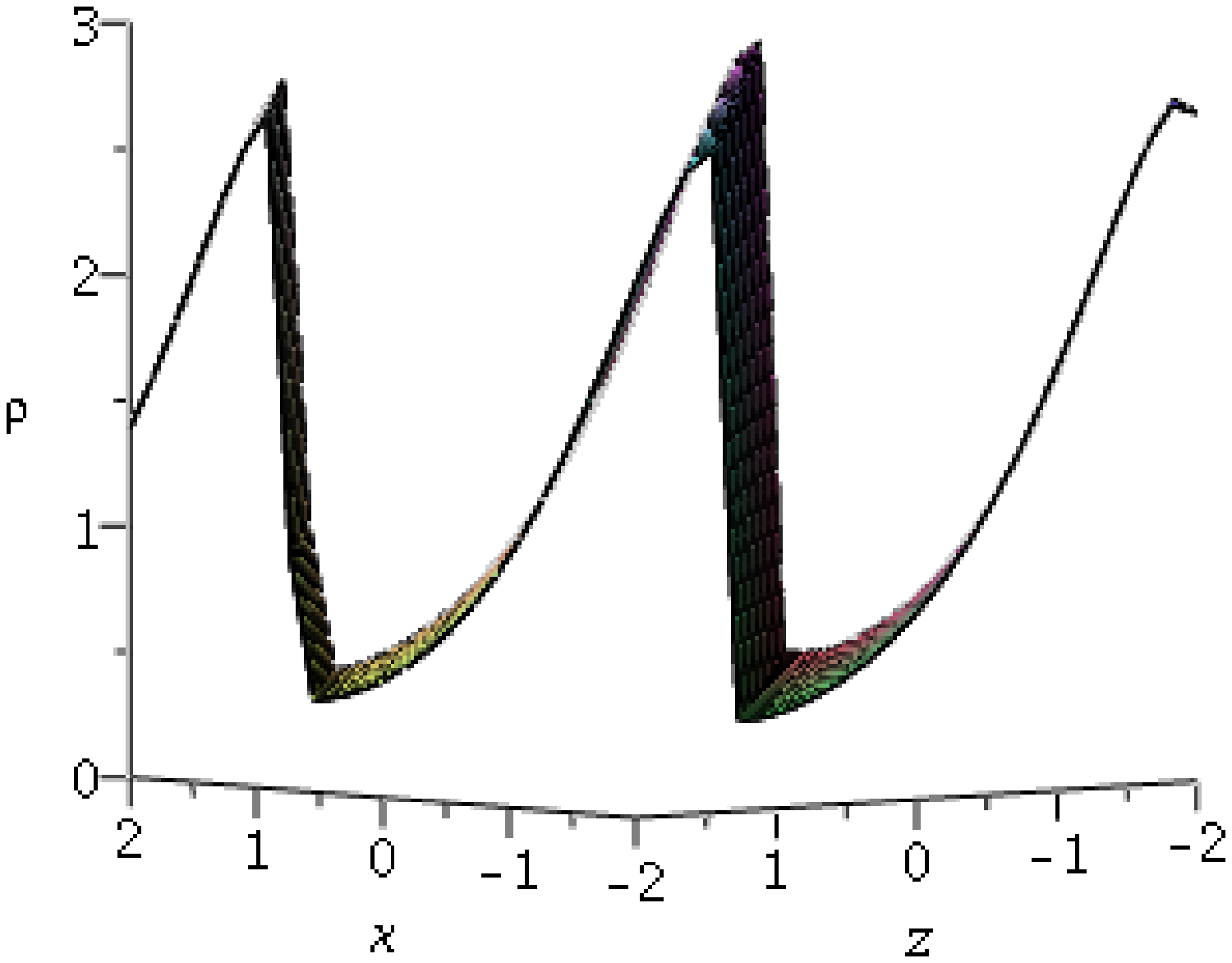}\\
\centering \includegraphics[width=6cm]{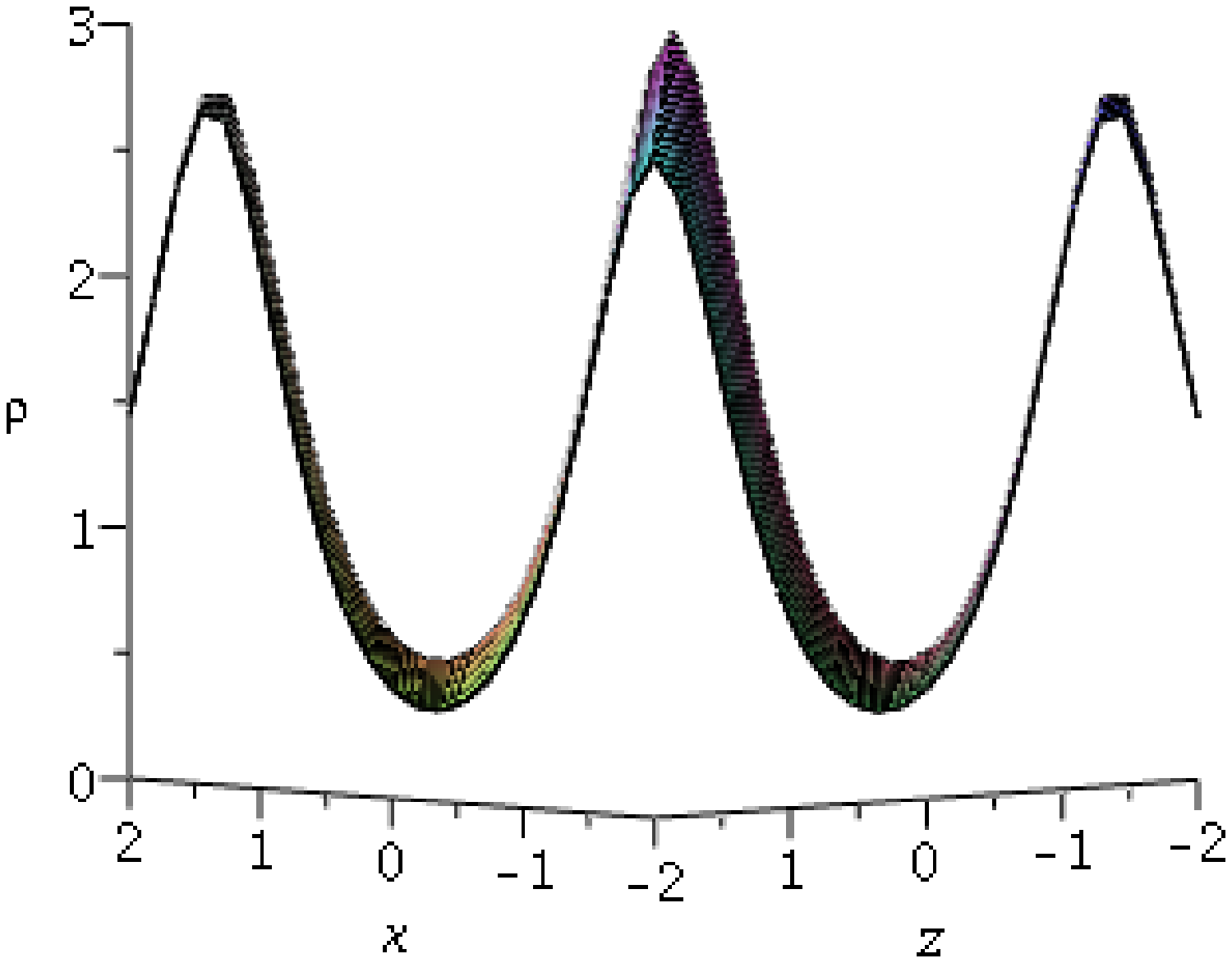} \includegraphics[width=6cm]{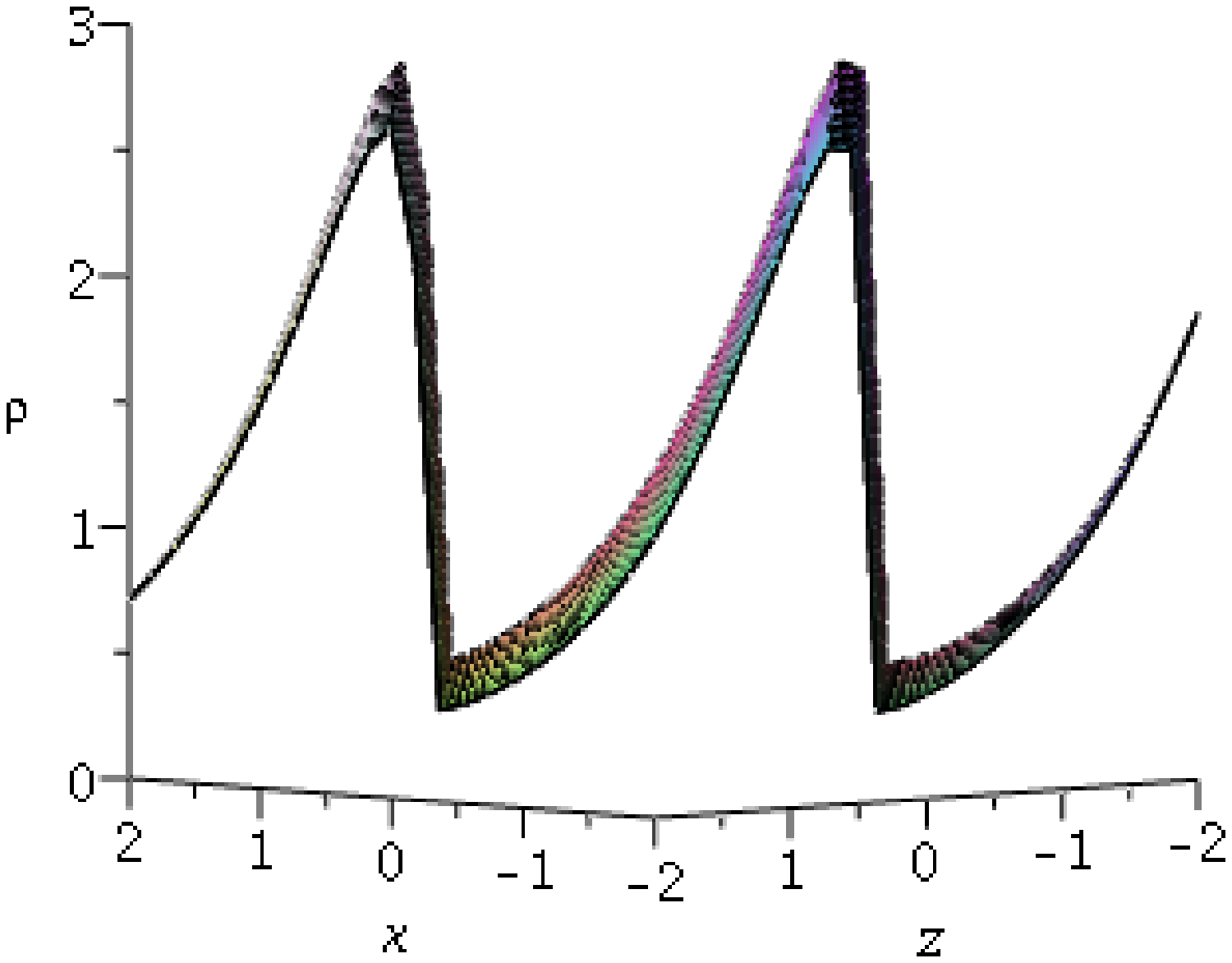}\\
\centering \includegraphics[width=6cm]{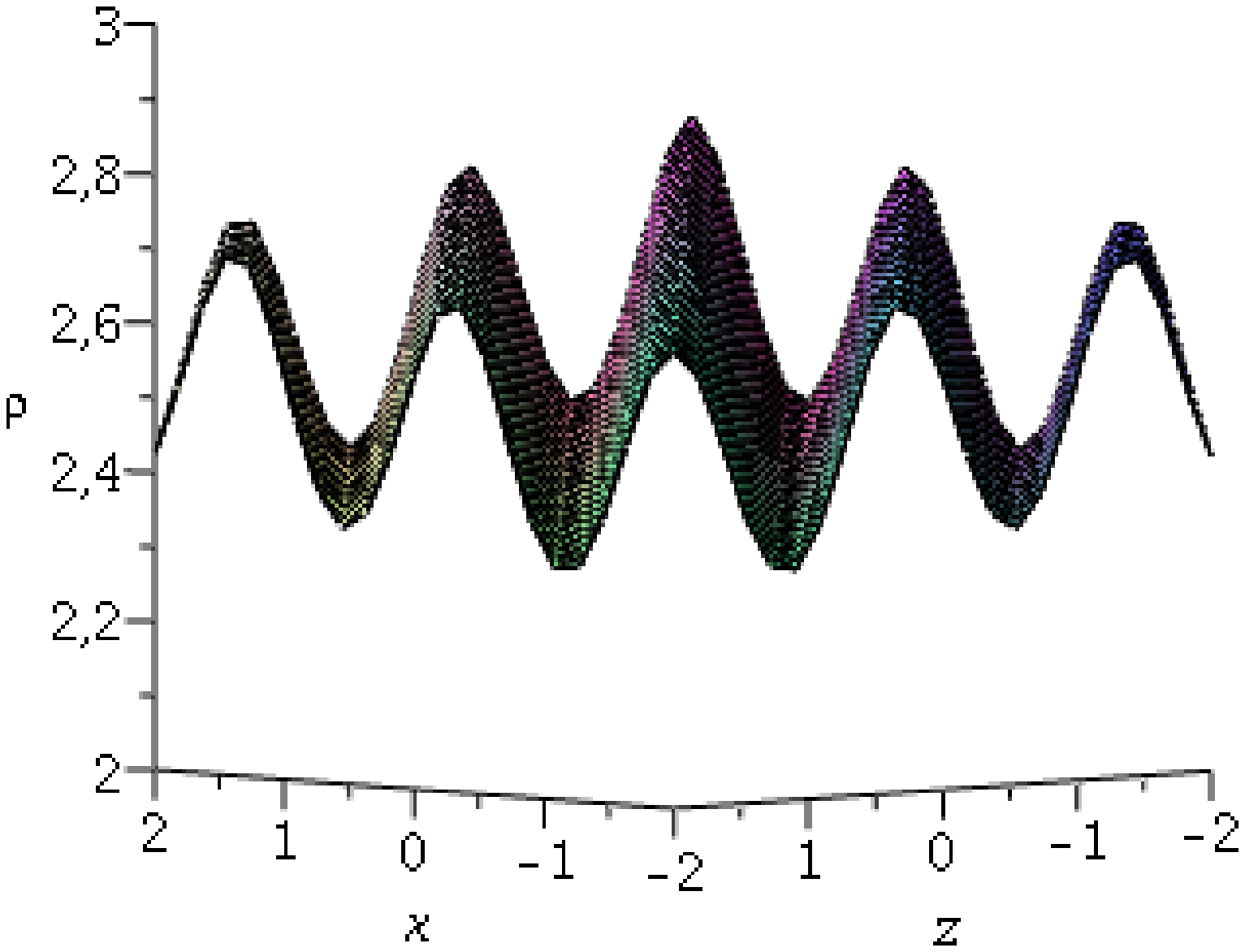} \includegraphics[width=6cm]{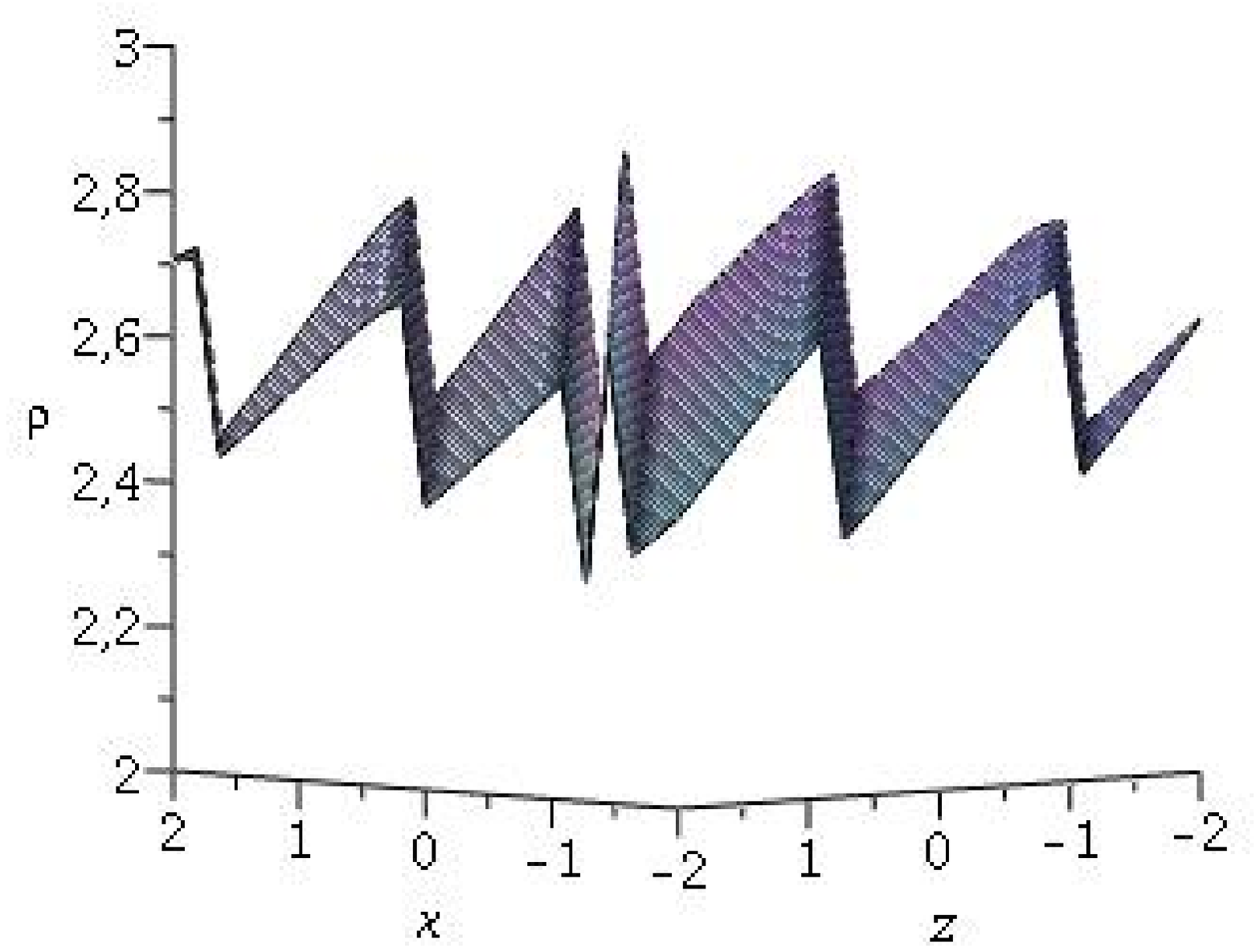}
\caption{Density distribution for the solution (\ref{E0Ae-sol}) at time $t=0$ and near the gradient catastrophe for the elliptic functions $B(r^1) = \mathrm{sn}(r^1,\frac{1}{2}),\mathrm{cn}(r^1,\frac{1}{2}) ,\mathrm{dn}(r^1,\frac{1}{2})$ .}
\label{fig-Density-dist}
\end{figure}

{\bf Subsystem $\{E^0, A_{+}\}$ :}
We look for solutions of the form (\ref{rank-k-sol}) defined by
\begin{equation*}
\begin{split}
&\vec{v}=\vec{v}(r^0,r^1), \quad \rho = \rho(r^0,r^1), \quad p = p(r^0,r^1), \\
&r^0 = - \left(\vec{v}\cdot\vec{g}\right) t + (g_1 + \Omega_3 v_2 - \Omega_2 v_3) x + (g_2 + \Omega_1 v_3 - \Omega_3 v_1) y + (g_3 + \Omega_2 v_1 - \Omega_1 v_2) z \\
&r^1 = - \left(\vec{\lambda}^1 \cdot \vec{v} + \left(\frac{\kappa p}{\rho} \right)^{1/2} |\vec{\lambda}^1| \right) t + \lambda^1_1 x + \lambda^1_2 y + \lambda^1_3 z, \quad \vec{\lambda}^1 = ( \lambda^1_1, \lambda^1_2, \lambda^1_3),
\end{split}
\end{equation*}
where the $\lambda^1_j$ are allowed to depend on the unknown functions.  The solution in this case exists only when $\kappa = 1$.  Following the same process as with subsystem $\{E_1^0, E\}$, the rank-$2$ solution exists when $\vec{g} \cdot \vec{\Omega} = 0$ and is given by
\begin{equation}
\label{E0Ae-sol}
\begin{split}
&v_1 = C_1 - \frac{\sqrt{A}}{|\vec{\Omega}|} \Omega_1 B(r^1), \\
&v_2 = \frac{1}{\Omega_1 \Omega_3} \left[ c_1 \Omega_1 - g_1 \Omega_1 + \Omega_2 \left( c_2 - g_2 + \Omega_3 \left( C_1 - \frac{\sqrt{A}}{|\vec{\Omega}|} \Omega_1 B(r^1) \right) \right)\right],\\
&v_3 = \frac{1}{\Omega_1} \left[ c_2 - g_2 + \Omega_3 \left( C_1 - \frac{\sqrt{A}}{|\vec{\Omega}|} \Omega_1 B(r^1) \right) \right],\\
&\rho = e^{r^0 / A + B(r^1)}, \quad p = A e^{r^0 / A + B(r^1)}, \quad A \in \mathbb{R}^+, C_1 \in \mathbb{R}.
\end{split}
\end{equation}
Here $B(r^1)$ is an arbitrary function, the Riemann invariants are given by
\begin{equation}
\label{E0Ae-RI}
\begin{split}
&r^0 = \frac{c_2 g_1 - c_1 g_2}{\Omega_3} t + c_1 x + c_2 y - \frac{c_1 \Omega_1 + c_2 \Omega_2}{\Omega_3} z,\\
&r^1 = \left(\frac{\sqrt{A}}{|\vec{\Omega}|} (\vec{\lambda}^1 \cdot \vec{\Omega}) B(r^1) - \lambda^1_1 C_1  + \sqrt{A} |\vec{\lambda}^1|\right.\\
& \left.  \quad + \frac{ \lambda^1_2 \Omega_1 (g_1 - c_1) - (\lambda^1_2 \Omega_2 + \lambda^1_3 \Omega_3)(c_2 - g_2 + C_1 \Omega_3)}{\Omega_1 \Omega_3} \right) t + \lambda^1_1 x + \lambda^1_2 y + \lambda^1_3 z,
\end{split}
\end{equation}
and $\vec{\lambda}^1$ is colinear with $\vec{\Omega}$, $\vec{\lambda}^1 \times \vec{\Omega} = 0.$ Note that this solution admits the gradient catastrophe for a certain time $T_0$ that we estimate by considering a first-order taylor expansion for $B(r^1) \sim B_0 + B_1 r^1$.  We obtain that the first-order partial derivative of (\ref{E0Ae-RI}) with respect to $t$ goes to infinity at time 
\begin{equation}
\label{E0Ae-CG-1}
T_0 = \frac{|\vec{\Omega}|}{B_1 \sqrt{A} \left( \vec{\lambda}^1 \cdot \vec{\Omega} \right) } .
\end{equation}
A physically interesting solution can be built by selecting $B(r^1)$ as the Jacobi elliptic function $\mathrm{sn}{(r^1,m)}$, where $m \in [0,1]$ is the modulus.  For this choice, solution (\ref{E0Ae-sol}), becomes 
\begin{equation}
\label{E0Ae-sol-sn}
\begin{split}
&v_1 = C_1 - \frac{\sqrt{A}}{|\vec{\Omega}|} \Omega_1\, \mathrm{sn}{(r^1,k)}, \\
&v_2 = \frac{1}{\Omega_1 \Omega_3} \left[ c_1 \Omega_1 - g_1 \Omega_1 + \Omega_2 \left( c_2 - g_2 + \Omega_3 \left( C_1 - \frac{\sqrt{A}}{|\vec{\Omega}|} \Omega_1\, \mathrm{sn}{(r^1,m)} \right) \right)\right],\\
&v_3 = \frac{1}{\Omega_1} \left[ c_2 - g_2 + \Omega_3 \left( C_1 - \frac{\sqrt{A}}{|\vec{\Omega}|} \Omega_1 \, \mathrm{sn}{(r^1,m)} \right) \right],\\
&\rho = e^{r^0 / A + \mathrm{sn}{(r^1,m)}}, \quad p = A e^{r^0 / A + \mathrm{sn}{(r^1,m)}}, \quad A \in \mathbb{R}^+, C_1 \in \mathbb{R},
\end{split}
\end{equation}
while the invariants adopt the implicit form
\begin{equation}
\label{E0Ae-RI-sn}
\begin{split}
&r^0 = \frac{c_2 g_1 - c_1 g_2}{\Omega_3} t + c_1 x + c_2 y - \frac{c_1 \Omega_1 + c_2 \Omega_2}{\Omega_3} z,\\
&r^1 = \left(\frac{\sqrt{A}}{|\vec{\Omega}|} (\vec{\lambda}^1 \cdot \vec{\Omega}) \mathrm{sn}{(r^1,m)} - \lambda^1_1 C_1  + \sqrt{A} |\vec{\lambda}^1|\right.\\
& \left.  \quad + \frac{ \lambda^1_2 \Omega_1 (g_1 - c_1) - (\lambda^1_2 \Omega_2 + \lambda^1_3 \Omega_3)(c_2 - g_2 + C_1 \Omega_3)}{\Omega_1 \Omega_3} \right) t + \lambda^1_1 x + \lambda^1_2 y + \lambda^1_3 z.
\end{split}
\end{equation}
Figure \ref{fig-Density-dist} illustrates the evolution of the density function $\rho(r^0,r^1)$ in (\ref{E0Ae-sol}) near the gradient catastrophe when the function $B(r^1)$ is chosen to be one the Jacobi elliptic functions $\mathrm{sn}(r^1,1/2),\mathrm{cn}(r^1,1/2),\mathrm{dn}(r^1,1/2)$ for a certain choice of the parameters. It should be noted that for solution (\ref{E0Ae-sol-sn}), the gradient catastrophe can be estimated to take place, according to formula (\ref{E0Ae-CG-1}), around the time 
$$T_0 = \frac{|\vec{\Omega}|}{\sqrt{A} (\vec{\lambda}^1 \cdot \vec{\Omega})}.$$

{\bf Subsystem $\{A_{\varepsilon}^0, E \}$ : } 
We now look for solutions of the form (\ref{rank-k-sol}) defined by
\begin{equation*}
\begin{split}
&\vec{v}=\vec{v}(r^0,r^1), \quad \rho = \rho(r^0,r^1), \quad p = p(r^0,r^1), \\
&r^0 = -\left(\vec{\lambda}^0 \cdot \vec{v} + \varepsilon \left(\frac{\kappa p}{\rho} \right)^{1/2} |\vec{\lambda}^0| \right) t + \lambda^0_1 x + \lambda^0_2 y + \lambda^0_3 z, \\
&r^1 = - \left(\vec{\lambda}^1 \cdot \vec{v} \right) t + \lambda^1_1 x + \lambda^1_2 y + \lambda^1_3 z, \quad \vec{\lambda}^s = ( \lambda^s_1, \lambda^s_2, \lambda^s_3), \quad s=0,1,
\end{split}
\end{equation*}
where the $\lambda^1_j$ are allowed to depend on the unknown functions.  In this case, the solution exists only when $\vec{\lambda}^0 = \varepsilon_1 |\vec{\lambda}^0| \vec{\Omega}$, $|\vec{\Omega}| = 1$ and $\vec{g} \cdot \vec{\Omega} = 0$.  Under these circumstances, the rank-$2$ solution takes the form
\begin{equation*}
\label{Case-3-sol-v3}
\begin{split}
&p = p_0, \quad \rho = \rho_0, \quad p_0, \rho_0 \in \mathbb{R},\\
&v_3 = \frac{1}{\Omega_3 \varepsilon_1} \left[ \varepsilon \sqrt{p_0 \kappa / \rho} - \varepsilon_1 ( \Omega_1  v_1 + \Omega_2 v_2) - c_0 \right],
\end{split}
\end{equation*}
where the functions $v_1, v_2$ satisfy the equations
\begin{equation}
\label{Case-3-eqs-v1-v2}
\left(\begin{array}{c} v_1 \\ v_2 \end{array} \right)_{r^0} = 
\left( \begin{array}{cc} d_1 & d_2 \\ d_3 & d_4 \end{array} \right) \left(\begin{array}{c} v_1 \\ v_2 \end{array} \right) + \left(\begin{array}{c} d_5 \\ d_6 \end{array} \right),
\end{equation}
with the constants being given by
\begin{eqnarray*}
&&\hspace{-2.5cm}d_1 = \frac{\Omega_1 \Omega_2}{\varepsilon \Omega_3 \sqrt{p_0 \kappa / \rho_0}}, 
\quad d_2 = \frac{{\Omega_2}^2 + {\Omega_3}^2}{\varepsilon \Omega_3  \sqrt{p_0 \kappa | / \rho_0}}, 
\quad d_3 = \frac{\Omega_2 \left(c_0 - \varepsilon \sqrt{p_0 \kappa / \rho_0 }\right) + g_1 \Omega_3 \varepsilon_1} {\varepsilon \Omega_3 \sqrt{p_0 \kappa / \rho_0}}, \\
&&\hspace{-2.5cm}d_4 =  - \frac{{\Omega_1}^2 + {\Omega_3}^2}{\varepsilon \Omega_3 \sqrt{p_0 \kappa  / \rho_0}}, 
\quad d_5 = - \frac{\Omega_1 \Omega_2}{\varepsilon \Omega_3 \sqrt{p_0 \kappa / \rho_0}},
\quad d_6 = \frac{ - \Omega_1 \left(c_0 - \varepsilon \sqrt{p_0 \kappa/\rho_0 }\right) + g_2 \Omega_3 \varepsilon_1} {\varepsilon \Omega_3 \sqrt{p_0 \kappa  / \rho_0}}. 
\end{eqnarray*}
The eigenvalues $\mu_{\pm}$ and eigenvectors $V_{\pm}$ of matrix $\left(\begin{array}{cc} d_1 & d_2 \\ d_3 & d_4 \end{array}\right)$ are given by
\begin{equation*}
\mu_{\pm} = \pm \varepsilon i \sqrt{\frac{\rho_0}{p_0 \kappa}}, \quad V_{\pm} = \left( \begin{array}{c} \frac{{\Omega_2}^2 + {\Omega_3}^2}{\Omega_1 \Omega_2 \mp i \Omega_3 } \\  -1 \end{array} \right).
\end{equation*}
The solution of equations (\ref{Case-3-eqs-v1-v2}) is then given by
\begin{equation*}
\label{Case-3-sol-v1-v2}
\begin{split}
&v_1 = F_1 (r^1) \cos{\left(\varepsilon \sqrt{\rho_0/\kappa p_0} r^0\right)} + F_2(r^1) \sin{\left(\varepsilon \sqrt{\rho_0/\kappa p_0} r^0\right)} + c_1,\\
&v_2 = -\frac{\Omega_1 \Omega_2}{\Omega_2^2 + \Omega_3^2} \left(F_1(r^1) \cos{\left(\varepsilon \sqrt{\rho_0/\kappa p_0} r^0\right)} + F_2(r^1) \sin{\left(\varepsilon \sqrt{\rho_0/\kappa p_0} r^0\right) + c_1} \right) \\
&+ \frac{\varepsilon_1 \Omega_3 }{\varepsilon(\Omega_2^2 + \Omega_3^2)} \left[ F_2(r^1) \cos{\left(\varepsilon \sqrt{\rho_0/\kappa p_0} r^0\right)} - F_1(r^1) \sin{\left(\varepsilon \sqrt{\rho_0/\kappa p_0} r^0\right)}\right]\\
& + \frac{\varepsilon \Omega_2 \sqrt{p_0 \kappa  / \rho_0} - c_0 \Omega_2 - \varepsilon_1 g_1 \Omega_2 \Omega_3}{\varepsilon_1 (\Omega_2^2 + \Omega_3^2)},
\end{split}
\end{equation*}
where $c_0 \in \mathbb{R}$, $F_1(r^1)$, $F_2(r^1)$ are arbitrary functions of the Riemann invariant $r^1$ and $c_1 \in \mathbb{R}$ is given by
\begin{equation*}
c_1 = \frac{\varepsilon_1  (g_2 \Omega_3^2 + \Omega_2^2 (g_2 + g_1 \Omega_1)) - c_0 \Omega_1 \Omega_3 + \varepsilon \Omega_1 \Omega_3 \sqrt{p_0 \kappa /\rho_0}}{\varepsilon_1 \Omega_3}.
\end{equation*}
The Riemann invariants then adopt the explicit form 
\begin{equation*}
\begin{split}
&r^0 = c_0 t + \varepsilon_1 \left( \Omega_1 x + \Omega_2 y + \Omega_3 z \right),\\
&r^1 = \varepsilon_1 \left(\varepsilon \sqrt{p_0 \kappa / \rho_0} -c_0 \right)t - \Omega_1 x - \Omega_2 y - \Omega_3 z.
\end{split}
\end{equation*}
This solution models the superposition of two traveling waves with the same direction but different phase velocities.

\begin{figure}[h]
\centering \includegraphics[width=7cm]{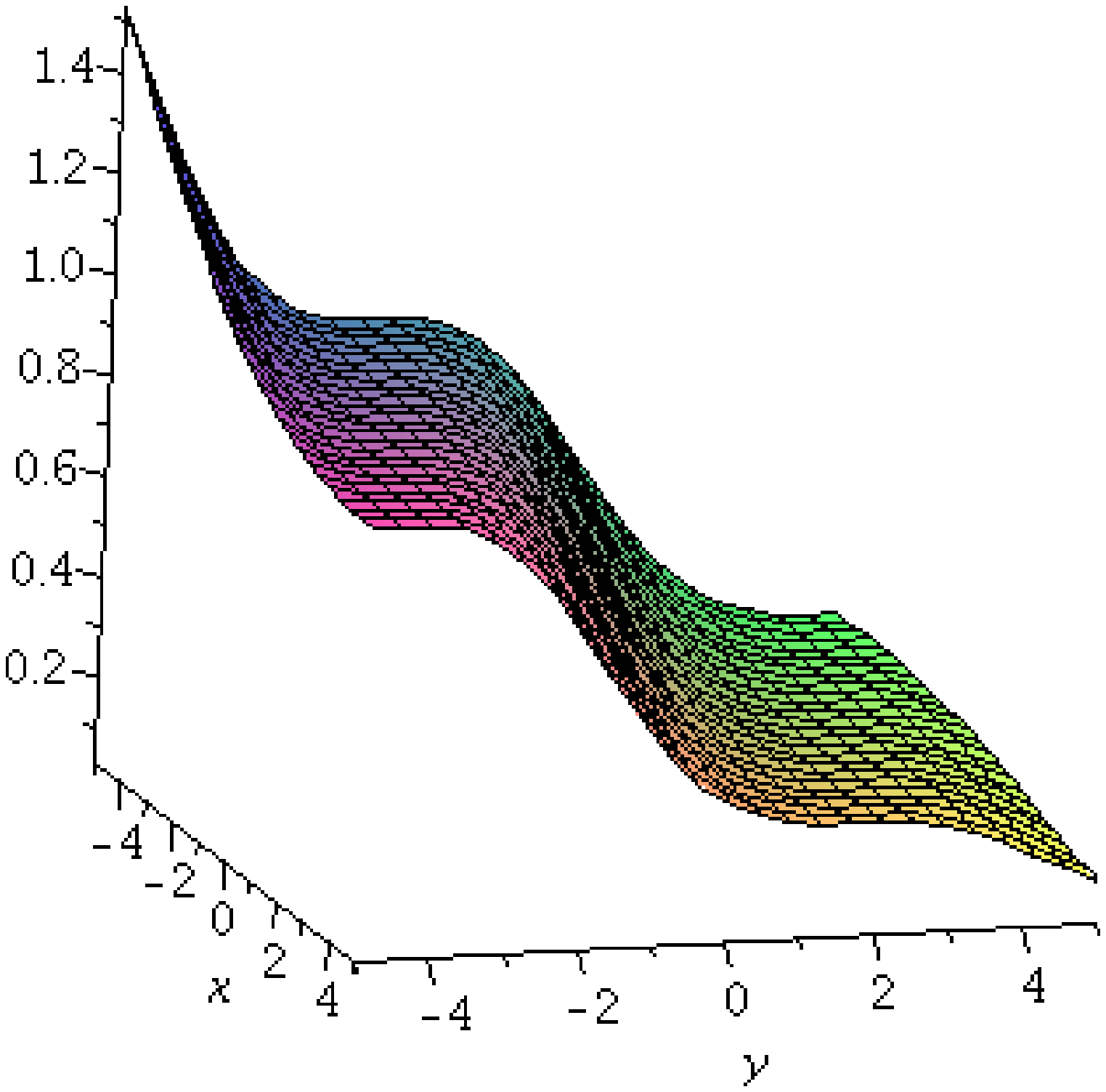} \includegraphics[width=7cm]{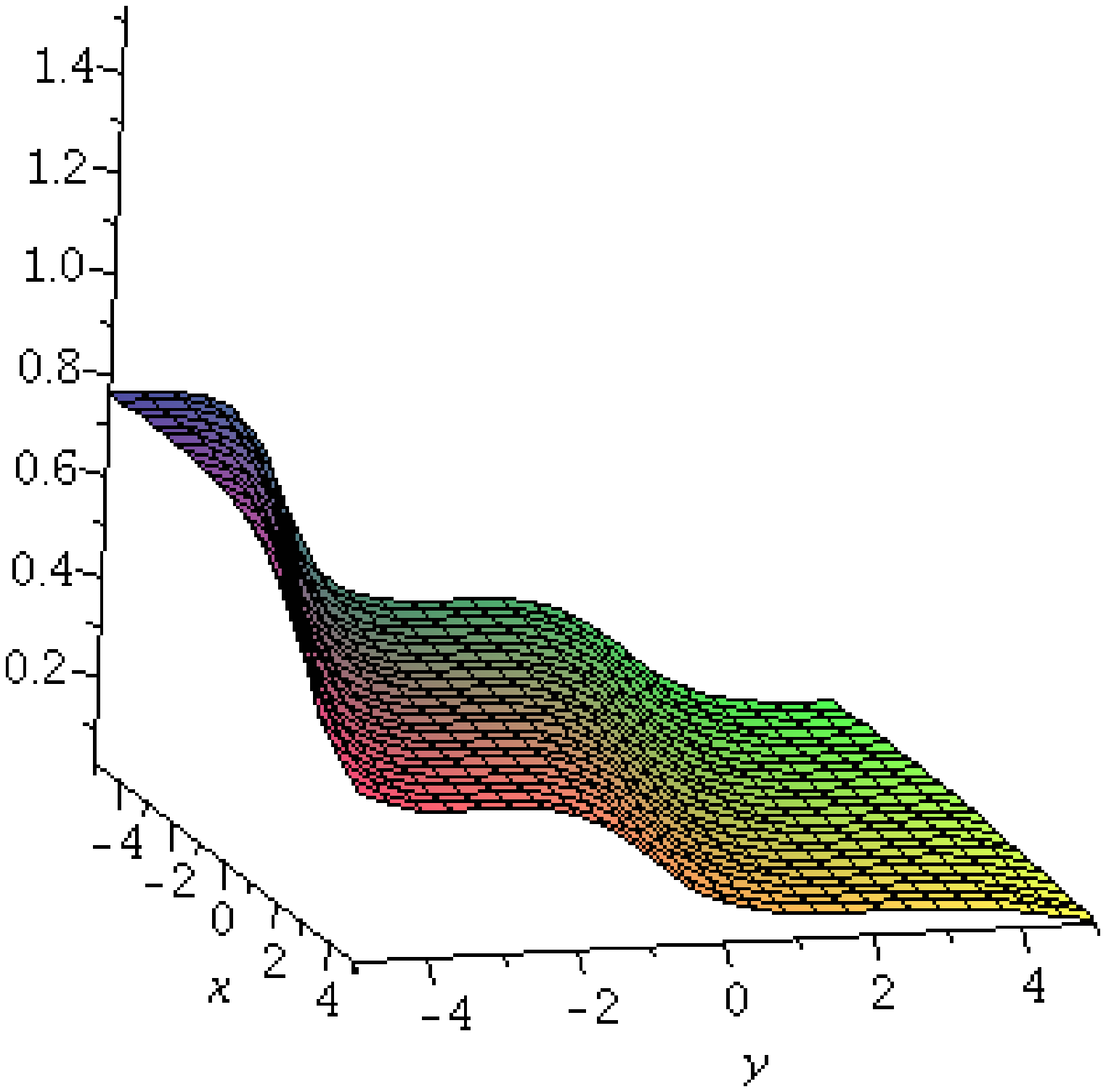}
\caption{Velocity vector field $v_2$ for solution (\ref{H0E-Lambert-sol}) at times $t=0$ and $t=1$ when $c_0 > 0$ and $F_1(r^1)=\sin{r^1}$.}
\label{fig:Lambert}
\end{figure}

{\bf Subsystem $\{H^0 E\}$} : We finally consider the case of the propagation of an entropic wave on an hydrodynamic state.  Since the required computations are very involved, we first assume, without loss of generality, that the coordinate system is chosen in such a way that the vector $\vec{\Omega}$ is given by $(0,0,1)$.  Using equations (\ref{rank-2-trace-eqs}), one can show that a laminar solution of equations (\ref{Euler-rotational}) exists when the wave vectors $\vec{\lambda}^0$ and $\vec{\lambda}^1$ are colinear and is given by
\begin{equation}
\label{H0E-Lambert-sol}
\begin{split}
&v_1 = g_2 + \frac{c_1}{c_2} \exp{ \left[ -W\left( \frac{e^s}{c_2 \left( c_0 + c_1 g_2 - c_2 g_1 \right)} \right) + s \right] }, \\
&v_2 = \exp{ \left[- W \left( \frac{e^s}{c_2 \left( c_0 + c_1 g_2 - c_2 g_1 \right)}\right) \right] } - g_1, \\
&v_3 = \int \frac{g_3}{c_0 + c_1 v_1 + c_2 v_2} \mathrm{d} r^0 + F_2(r^1), \\
&\rho = \rho(r^1) > 0, \quad p = p_0, \quad s = \frac{c_1 (r^0 + F_1(r^1)) - g_1}{c_2 \left( c_0 + c_1 g_2 - c_2 g_1 \right)},
\end{split}
\end{equation}
where the Riemann invariants satisfy the relations
\begin{equation}
\begin{split}
&r^0 = c_0 t + c_1 x + c_2 y, \quad c_2 \neq 0, \quad {c_1}^2 + {c_2}^2 = 1, \quad \vec{\Omega} = (0,0,1), \\
&r^1 = - \left( c_1 g_2 - c_2 g_1 + \exp{\left[ - W \left( \frac{e^s}{c_2 \left( c_0 + c_1 g_2 - c_2 g_1 \right)} \right)\right] }\right) t + c_1 x + c_2 y.
\end{split}
\end{equation}
Here, the functions $F_1(r^1)$, $F_2(r^1)$ are arbitrary functions and $\rho(r^1)$ is an arbitrary positive function of the Riemann invariant $r^1$. The function $W(\phi)$ stands for the principal branch of the Lambert $W$ function which is defined by the relation
\begin{equation}
W\left(\phi \right) \exp{\left[ W\left(\phi\right)\right] } = \phi.
\end{equation}
The Lambert function $W(\phi)$ is positive and increasing when its argument is greater than $0$ and negative only for $-e^{-1} < \phi < 0$. Moreover, it is unbounded when $\phi$ tends to $+\infty$. Hence, the velocity components $v_1, v_2$ are bounded for positive values of 
$$\phi = \frac{e^s}{c_2(c_0 + c_1 g_2 - c_2 g_1)}.$$  
Also, the solution is perturbed in accordance with the nature of the arbitrary function $F_1(r^1)$. This situation is illustrated in figure \ref{fig:Lambert} for  the case of a periodic perturbation, $F_1(r^1) = \sin{r^1}$.


\section{Final remarks}


The main result of this paper consists of extending the Riemann invariant method to inhomogeneous quasilinear systems with coefficients depending only on dependent variables $u$.  The conditional symmetry method, together with the Cayley-Hamilton idea, was adapted to partial differential equations in such a way as to allow the applicability of the Riemann invariants method.  The conditions (\ref{rank-2-trace-eqs}) and (\ref{rank-3-trace-eqs}) are necessary and sufficient for the existence of rank $2$ and $3$ solutions of the system (\ref{original-system}), respectively.  In the proof of theorems $1$ and $2$, we obtain that the gradient of the matrix functions $M_1^{-1}$ or $M_2^{-1}$ becomes infinite for certain values of $x^0$ (the solution and its derivatives do not remain bounded).  This means that we are dealing with a gradient catastrophe and some discontinuities can arise such as, for example, shock waves.  Thus, the existence of Riemann invariants implies that it is not possible to introduce the concept of a weak solution in the broad sense.

These theoretical considerations have been illustrated by the examples of inhomogeneous equations of fluid dynamics in the presence of gravitational and Coriolis external forces.  We were able to construct new classes of rank-$2$ solutions expressible in terms of Riemann invariants in the cases $E^0 E, E^0 A_{\varepsilon}, A^0_{\varepsilon} E, H^0 E, H^0 A_{\varepsilon}$.  Some of these solutions admits a gradient catastrophe.  This results from the fact that the first derivatives of Riemann invariants become infinite after some finite time.  These solutions, in their general form, possess some degree of freedom, that is, depend on one or two arbitrary functions of one variable (Riemann invariant) depending on the case.  This arbitrariness allows one to change the geometrical properties of the governed fluid flow in such a way as to displace the singularities to a sufficient extent.  This fact is of some significance since even for arbitrary smooth and sufficiently small initial data at $t=t_0$, the magnitude of the first derivatives of the Riemann invariants become unbounded in some finite time $T$.  This time can be estimated for each solution. \\

\noindent {\bf Acknowledgement}\\
This work has been supported by a research fellowship from NSERC of Canada.

\bibliographystyle{plain}
\bibliography{Biblio}

\end{document}